# Solid-state neutron detectors based on thickness scalable hexagonal boron nitride

K. Ahmed,[1,a)] R. Dahal,[1,] A. Weltz,[2] James J.-Q. Lu,[1] Y. Danon,[2] and I. B. Bhat[1]

[1]*Department of Electrical, Computer, and Systems Engineering, Rensselaer Polytechnic Institute, Troy, NY 12180, USA*
[2]*Department of Mechanical, Aerospace and Nuclear Engineering, Rensselaer Polytechnic Institute, Troy, NY 12180, USA*

(Dated: October 11, 2016)

This paper reports on the device processing and characterization of hexagonal boron nitride (hBN) based solid-state thermal neutron detectors, where hBN thickness varied from 2.5 to 15 µm. These natural hBN epilayers (with 19.9% $^{10}$B) were grown by a low pressure chemical vapor deposition process. Complete dry processing was adopted for the fabrication of these metal-semiconductor-metal (MSM) configuration detectors. These detectors showed intrinsic thermal neutron detection efficiency values of 0.86%, 2.4%, 3.15%, and 4.71% for natural hBN thickness values of 2.5, 7.5, 10, and 15 µm, respectively. Measured efficiencies are very close ($\geq$ 92%) to the theoretical maximum efficiencies for corresponding hBN thickness values for these detectors. This clearly shows the hBN thickness scalability of these detectors. A 15 µm thick hBN based MSM detector is expected to yield an efficiency of 21.4%, if enriched hBN (with ~100% $^{10}$B) is used instead of natural hBN. These results demonstrate that the fabrication of hBN thickness scalable highly efficient thermal neutron detectors is possible.

High efficiency neutron detectors are essential for homeland security and nuclear safeguards, since neutrons are a very specific indicator of special nuclear materials (SNMs). Development of solid-state neutron detectors (SSNDs) represents an emerging area of research, because the existing highest efficiency $^3$He gas based neutron detectors have drawbacks such as high cost, bulky nature, and high pressure and high bias voltage requirement [1-3]. SSNDs utilize high thermal neutron capture cross section ($\sigma$) values of neutron sensitive isotopes such as $^{10}$B ($\sigma$ = 3840 barns) and $^6$Li ($\sigma$ = 940 barns) [2]. Most SSNDs are microstructured Si filled with $^{10}$B or $^6$Li, and are of heterogeneous type with separate neutron conversion and charge collection regions. $^{10}$B has a natural abundance of ~19.9% [2] and is a constituent element of hexagonal boron nitride (hBN). hBN is therefore an excellent material for homogenous SSND fabrication, where both neutron conversion and charge collection can occur in this semiconductor. Large bandgap of hBN (5.5 eV to 6 eV [4-6]) makes hBN SSNDs radiation hardened. hBN SSNDs are also relatively insensitive to gamma rays [7]. The theoretical requirement of hBN thickness (assuming 100% $^{10}$B) is 35 µm and 80 µm, for the detection efficiency of 50% and 80%, respectively, as indicated by our simulation [7]. This work reports on the growth of thick hBN of thickness ranging from 2.5 to 15 µm using a chemical vapor deposition (CVD) process. SSNDs with metal-semiconductor-metal (MSM) configuration were fabricated with these hBN films. Charge transport in these devices is along a-axis, which represents the highest charge carrier mobility path in hBN [8]. Device capacitance ($C_{dev}$) and dark current ($I_{dev}$) of an hBN MSM SSND are orders of magnitude lower compared to microstructured Si based SSNDs [3, 9]. Since lower values of $C_{dev}$ and $I_{dev}$ allow for scaling to larger device areas, hBN MSM SSNDs can be scaled to much larger detection area compared to microstructured Si based SSNDs [3, 9]. MSM devices with 2.5 to 15 µm thick hBN yielded close ($\geq$ 92%) to theoretically expected detection efficiencies, demonstrating hBN thickness scalability of these devices.

A low pressure CVD system was employed for the epitaxial growth of hBN on sapphire substrates. SiC coated graphite susceptors were used, which were heated by an induction heating system. Precursors were triethylboron (TEB) and ammonia (NH$_3$) for B and N, respectively. High purity Hydrogen (H$_2$) was used as the carrier gas with a fixed flow of 2 SLM for the entire growth durations. 100 Torr was the chamber pressure. In order to reduce the large lattice mismatch between hBN and sapphire, a 10-minute low temperature (850 °C) sapphire nitridation step was introduced before hBN growth [5, 9]. The amorphous and thin nitridated layer works as a nucleation layer for hBN and effectively reduces the lattice mismatch mentioned earlier [5, 9]. hBN growth was performed at a temperature of 1350 °C. 2.5 µm and 15 µm thick hBN films exhibited broad (002) hBN peaks at 26.2° and 26°, respectively [8, 9], indicating slightly larger c-lattice constant compared to that of bulk hBN (6.66 Å [10, 11]). Figure 1 (a) shows the schematic layer structure of the grown hBN films. The optical image of an 8 µm thick hBN film grown on a 2-inch diameter double side polished sapphire wafer is illustrated in Fig. 1 (b).

hBN SSNDs with an MSM architecture were fabricated with a complete dry processing technique developed at RPI. hBN films grown on sapphire substrates peel off the substrates as soon as they come in contact with any liquid if the film thickness is > 0.3 µm. Hence, wet processing (e.g., photolithography) was not an option for the fabrication of hBN MSM SSNDs. For the dry processing, stainless steel shadow masks were used for both hBN etching (to define the metal electrode areas) and metal deposition (to form the

a) Electronic mail: ahmedk2@rpi.edu.

electrodes). hBN etching was done by inductively coupled plasma-reactive ion etching (ICP- RIE) using $SF_6$ plasma, and the etch rate was ~ 0.8 µm/min. E-beam evaporation was used to deposit metal for electrode formation. Metal contacts were 30 nm Ni followed by 250 nm Ti. Four MSM devices were fabricated and characterized. The devices are: device A (hBN thickness, t = 2.5 µm, electrode spacing, L= 1 mm, and area, A = 40 $mm^2$), device B (t =7.5 µm, L= 250 µm, and A = 5 $mm^2$), device C (t =10 µm, L= 200 µm, and A = 11.2 $mm^2$), and device D (t =15 µm, L= 250 µm, and A = 14 $mm^2$). The optical image of device D is shown in Fig. 1 (c).

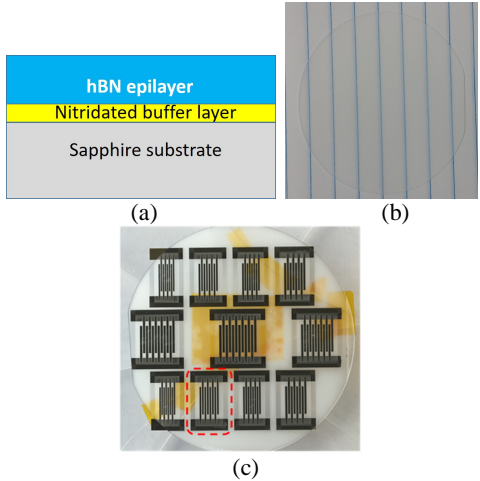

Fig. 1. (a) Schematic of the layer structure of an hBN epilayer grown on sapphire, (b) optical image of an 8 µm hBN film on a 2-inch double side polished sapphire wafer placed on a ruled paper, and (c) optical image of an array of detectors fabricated using a 15 µm thick hBN film. Detector D (with L= 250 µm and A = 14 $mm^2$) is the one highlighted by a red dashed rectangle. The yellow portion is the double sided Kapton tape at the backside of sapphire wafer used to attach it to a glass plate.

hBN SSNDs (with t ranging from 2.5 to 15 µm) showed strong response to deep UV light, as evidenced from the I-V characteristics of the detector D (with t = 15 µm) shown in Fig. 2 (a). Undoped hBN epilayers showed electrical resistivity of ~$3 \times 10^9$ Ω cm, which guarantees a very low dark current for the fabricated MSM devices. No persistent photoconductivity (PPC) effects are present in the detectors, as indicated by the photocurrent decay kinetics for the devices at a bias voltage of 150 V (not shown here). Figure 2 (b) shows the spectral response of the same hBN MSM detector at a bias voltage of 10 V. The detector showed peak responsivity wavelength of 215 nm and a sharp cutoff wavelength of 225 nm, which corresponds well with the indirect bandgap of 5.47 eV measured for hBN films [5].

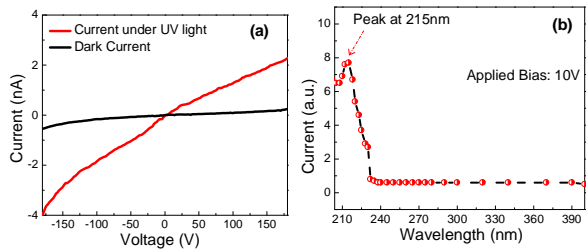

Fig. 2. (a) I-V characteristics and (b) spectral response of detector D. t = 15 µm and L = 250 µm for this detector.

Neutron response of the fabricated detectors was measured using a californium-252 ($^{252}$Cf) neutron source. Neutron capture cross section decreases with increasing neutron energy. Hence, it is desirable to detect thermal neutrons (0.0259 eV) using the neutron detectors. The $^{252}$Cf source emits fission neutrons and is encapsulated in a large high-density polyethylene (HDPE) block that acts as the moderating material. The moderator size is $61 \times 61 \times 40$ $cm^3$, and the $^{252}$Cf source is placed 2.5 cm away from the center of the $61 \times 61$ $cm^2$ front face [3, 12]. The moderator reduces the speed of fast neutrons and turns them into thermal neutrons. The detectors were enclosed in an aluminum box, which reduces the electronic noise by shielding the devices from external electromagnetic radiation. All the pulse height spectra measurements related to neutron response were performed using the following electronics: an ORTEC 142PC charge sensitive preamplifier, an ORTEC 672 pulse shaping amplifier, and an ORTEC ASPEC-927 multichannel analyzer (MCA). Monte Carlo N-Particle Transport Code (MCNP) simulation was performed previously [3] to calculate the neutron flux spectrum as a function of incident energy on the detectors. The neutron flux shape closely matched a Maxwellian spectrum, where the kT product is 0.034 eV [3]. Associated σ value for $^{10}$B at this kT value is determined to be 3450 barns (i.e., $3450 \times 10^{-24}$ $cm^2$), whereas σ equals 3840 barns at kT = 0.0259 eV. Σ (= σN) is the macroscopic cross section of $^{10}$B, where N is the atomic number density of $^{10}$B. Σ values in this work are 35.01 $cm^{-1}$ and 175.93 $cm^{-1}$ for a natural hBN layer (with 19.9% $^{10}$B) and an enriched hBN layer (with ~100% $^{10}$B), respectively. A comparison between the same MNCP calculations and experimentally measured neutron fluxes showed that overall neutron flux uncertainty is less than 5% [3].

Carrier mobility-lifetime product (µτ) is a crucial parameter for a neutron detector, since it determines the charge collection efficiency and hence the performance of the detector. Most of the electrons and holes generated by the nuclear reactions are collected by the electrodes under the applied electric field if the recombination time (τ) of the free carriers is larger than the carrier transit time ($\tau_t = L/\mu E = L^2/\mu V$) in the material, where L is the distance between two electrodes, V is the applied bias voltage, and E is the applied electric field. Therefore, µτ product should meet the requirement, $\mu\tau \geq L^2/V$ for a SSND [8, 13]. Higher µτ products of the electrons and holes enable collection of the carriers at a smaller electric field for a given electrode spacing. To obtain the µτ product for the carriers, neutron count rate (C) of device D (under thermal neutron irradiation from the moderated $^{252}$Cf source) was measured as a function of the applied voltage, which is shown in Fig. 3. The count rates were recorded above the background noise level. The count rate as a function of voltage is described by the Hecht's equation [14],

$$C(V) = C_0 \left[ \left\{ \frac{\mu_n \tau_n}{L^2} V - \frac{\mu_n \tau_n}{L^2} V\, e^{-\frac{L^2}{\mu_n \tau_n V}} \right\} + \left\{ \frac{\mu_p \tau_p}{L^2} V - \frac{\mu_p \tau_p}{L^2} V\, e^{-\frac{L^2}{\mu_p \tau_p V}} \right\} \right] \quad (1)$$

where $\mu_n \tau_n$ and $\mu_p \tau_p$ denote μτ products for electron and hole respectively. $C_0$ is the saturation count rate. Both $\mu_n \tau_n$ and $\mu_p \tau_p$ values are obtained from fitting the measured count rate vs. V curve with the relation of Eq. (1). Fitted μτ values for this 15 μm hBN film (of detector D) are, $\mu_n \tau_n = 2.8 \times 10^{-7}$ cm²/V and $\mu_p \tau_p = 3.2 \times 10^{-8}$ cm²/V.

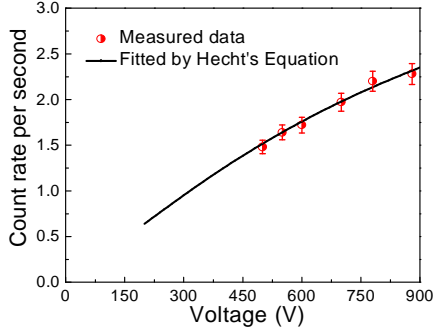

Fig. 3. Count rate vs. bias voltage of detector D under thermal neutron irradiation. Half-filled circles are the measured data and the solid line is the fitted data using Eq. (1). Detector D has t =15 μm, L= 250 μm, A = 14 mm². Error bars for the measured count rates are indicated in the plot.

Intrinsic thermal neutron detection efficiency (η) was measured for four different detectors (A, B, C, and D). For η measurement, the devices were placed 8 cm away from the front face of the moderator housing. Gold foil activation method was previously used to calibrate the thermal neutron flux (φ) at this position. Calculated φ values at this position were 380 n/cm²-s, 320 n/cm²-s, 300 n/cm²-s, and 298 n/cm²-s, respectively for the four devices at the time of the corresponding measurements. Applied bias voltages for the devices were 700 V, 450 V, 600 V, and 700 V, respectively. Three different measurements were taken to obtain η of each of the detectors. The first two measurements were done with the ²⁵²Cf source kept inside the moderator housing. This moderated neutron source emits not only thermal neutrons, but also some fast neutrons and gamma rays. The second measurement had a 2 mm thick cadmium (Cd) sheet in between the bare device and the moderator. Cd absorbs the thermal neutrons and allows the fast neutrons and gamma rays to pass through it. The difference in the count rates ($C_{Bare} - C_{Cd}$) represents the count rates caused by the thermal neutrons, where $C_{Bare}$ and $C_{Cd}$ are the count rates from the first and the second measurements, respectively. The ²⁵²Cf source was taken out of the moderator for the third measurement, which determined the electronic noise level. The equation for measuring η is as follows:

$$\eta = \frac{C_{Bare} - C_{Cd}}{\varphi A} \quad (2)$$

The measured efficiency for the devices (A, B, C, and D) are $0.86 \pm 0.03\%$, $2.4 \pm 0.2\%$, $3.15 \pm 0.12\%$, and $4.71 \pm 0.18\%$, respectively. Fig. 4 (a), (b), (c), and (d) illustrate the associated pulse height spectra, respectively. Neutron interaction probability (p) of an hBN layer can be estimated from the equation, $p = 1 - e^{-\Sigma t}$. The maximum theoretical intrinsic thermal neutron detection efficiency ($\eta_{theory}$) for an hBN device is equal to the corresponding p value. $\eta_{theory}$ values for these four natural hBN devices are 0.87%, 2.59%, 3.44%, and 5.11%, respectively. Effective neutron conversion efficiency ($\eta_{conv} = \eta/\eta_{theory}$) is termed as the fraction of the incident thermal neutrons that generated charge above the background noise level and therefore were counted by the detector. $\eta_{conv}$

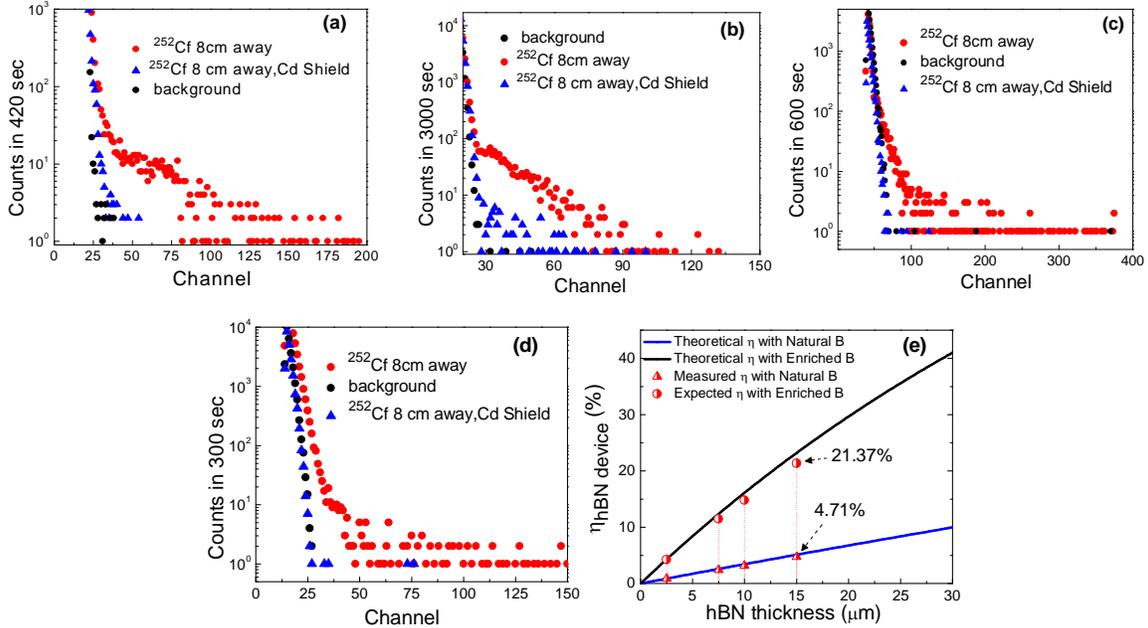

Fig. 4. Pulse height distribution of thermal neutrons measured with (a) detector A, (b) detector B, (c) detector C, and (d) detector D. (e) Theoretical and measured efficiency values of hBN detectors with different t values.

TABLE I. Efficiency values of SSNDs with hBN films of different thicknesses.

| Device | hBN thickness(μm) | Measured $\eta$, natural hBN | $\eta_{theory}$, natural hBN | $\eta_{conv}$ | Expected $\eta$, enriched hBN | $\eta_{theory}$, enriched hBN |
|---|---|---|---|---|---|---|
| A | 2.5 | 0.86% | 0.87% | 98% | 4.21% | 4.3% |
| B | 7.5 | 2.4% | 2.59% | 93% | 11.5% | 12.36% |
| C | 10 | 3.15% | 3.44% | 92% | 14.84% | 16.13% |
| D | 15 | 4.71% | 5.11% | 92% | 21.37% | 23.19% |

values for the devices are thus ≥ 92%. Natural TEB source (with 19.9% $^{10}$B) was used for hBN growth in this work. Using the enriched TEB source (with ~100% $^{10}$B) should not impact material properties of the hBN films, but increase the detection efficiency significantly. Considering the associated $\eta_{conv}$ values, expected efficiencies for the four devices with enriched hBN films are 4.21%, 11.5%, 14.84%, and 21.37%, respectively. These results are summarized in Table I. Achievement of $\eta_{conv} \geq 92\%$ for SSNDs with t ranged from 2.5 to 15 μm illustrates the thickness scalability for these detectors. Fig. 4 (e) demonstrates this thickness scaling phenomenon for hBN MSM SSNDs. If these results are extrapolated to a thicker (50 μm) hBN layer on sapphire with this type of MSM device structure, a natural hBN device (with 19.9% $^{10}$B) should achieve an efficiency close to 20% and an enriched hBN device (with ~100% $^{10}$B) should achieve an efficiency close to 60%. Further, low device capacitance (due to device geometry and hBN properties) and low dark current (arising from high resistivity of hBN) ensure that these hBN SSNDs can be scaled up to much larger areas, as compared to microstructured Si based SSNDs.

In summary, CVD grown hBN films were used to fabricate neutron detectors with an MSM device structure. Developed detector fabrication process is a complete dry one involving shadow masks, which solves the typical problem of thick hBN films peeling off the sapphire substrate during wet processing. This process enabled MSM device fabrication with hBN films of thickness of up to 15 μm, with the possibility of even thicker hBN based device fabrication. hBN MSM devices, with thickness ranging from 2.5 to 15 μm, showed close (≥ 92%) to the corresponding theoretical maximum detection efficiencies, indicating the hBN thickness scalability of these devices. Improvement in the material quality and the dry fabrication process and the usage of enriched TEB for hBN growth would further improve the performances of these detectors.

The authors acknowledge the support of the staff of the Rensselaer Polytechnic Institute Micro and Nano Fabrication Clean Room (RPI MNCR). This work is financially supported by the US Department of Homeland Security, Domestic Nuclear Detection Office, under the grants ECCS-1348269 and 2013-DN-077-ER001.